# Relation between seismicity and pre-earthquake electromagnetic emissions in terms of energy, information and entropy content.

S. M. Potirakis[1], G. Minadakis[2], and K. Eftaxias[3]

[1] Department of Electronics Engineering, Technological Education Institute (TEI) of Piraeus, 250 Thivon & P. Ralli, GR-12244, Aigaleo, Athens, Greece, tel.: +302105381550, e-mail: spoti@teipir.gr .

[2] Department of Electronic and Computer Engineering, Brunel University Uxbridge, Middlesex, UB8 3PH, U.K, tel.: +302103452834, e-mail: george.minadakis@brunel.ac.uk .

[3] Department of Physics, Section of Solid State Physics, University of Athens, Panepistimiopolis, GR-15784, Zografos, Athens, Greece, tel.: +302107276733, e-mail: ceftax@phys.uoa.gr .

**Abstract**

In this paper we show, in terms of Fisher information and approximate entropy, that the two strong impulsive kHz electromagnetic (EM) bursts recorder prior to the Athens EQ (7 September 1999, magnitude 5.9) present compatibility to the radar interferometry data and the seismic data analysis, which indicate that two fault segments were activated during Athens EQ. The calculated Fisher information and approximate entropy content ratios closely follow the radar interferometry result that the main fault segment was responsible for 80% of the total energy released, while the secondary fault segment for the remaining 20%. This experimental finding, which appears for the first time in the literature, further enhances the hypothesis for the seismogenic origin of the analyzed kHz EM bursts.

## 1      Introduction

The research related to short term earthquake (EQ) prediction does not appear to be seen favorably by the scientific community. There have been expressed arguments up to the extreme that it is impossible to have any precursory activity (Uyeda et al., 2009). These views do not appear to be unjustified if someone considers the difficulties coupled with the facts that the large EQs are rare, they are extremely complex phenomena, and that there is a great variety of possible preseismic patterns.

Although, many issues related to EQ generation are not fully clarified yet, it is reasonably expected that the preparatory process of EQ's has various aspects which may be observed before the final event. Therefore, the multi-disciplinary character of the science of EQ prediction is indubitable. In this work, we deal with this subject. Physical fields which are caused from fracture phenomena allow the real-time observation of the evolution of damage of materials under mechanical loading. Cracks' opening is producing electromagnetic (EM) emissions in a wide frequency band from kHz to MHz. Both laboratory and geophysical scale experimental data report the detection of these precursors. Monitoring techniques based on these fracture-induced fields are fundamental for the comprehension of fracture mechanism, as well as for the development of models of rock/focal area behavior. EM precursors have not been adequately accepted as valid yet. Two criteria have been recently proposed for the acknowledgment of an observed signal as EQ precursor (Cicerone et al., 2009). The first one was "the reported existence of credible scientific evidence for anomalies in the observables prior to at least some earthquakes" and the other was the existence of "acceptable physical models to explain the existence of the precursor". We will show that the EM precursors under study satisfy the aforementioned two criteria.

We focus on a well documented pre-seismic EM signal associated with the Athens EQ (7 September 1999, magnitude 5.9). Two strong impulsive kHz EM emissions were recorded, with a sampling rate of 1 sample/s. The anomaly came to an end 9 hours before the main event. The first strong emission lasted for 12 hours; 12 hours of background noise were recorded after that, followed by the second strong emission of 17 hours duration (Eftaxias et al., 2001; Papadimitriou et al., 2008). The existence of these kHz EM signals has been criticized in the past (Pham and Geller, 2002). However, a series of papers which followed, e.g., (Eftaxias et al., 2004; Kapiris et al., 2004; Contoyiannis et al., 2005; Karamanos et al., 2006; Papadimitriou et al., 2008; Eftaxias et al., 2008; Potirakis et al., 2012; Minadakis et al., 2012), have further supported the view that the detected anomalies were preseismic. It has been repeatedly clarified that such signals have been recorded only in the case of strong surface EQs with epicenters in land (or near coast-line), a fact which provides an explanation for their transmission to the air. Note that the recently proposed fractal geoantenna model (Eftaxias et al., 2004) justifies why there is an increased possibility of detection of such EM radiation, due to the increased radiated power compared to the power radiated in the case of a dipole model.

In terms of seismic energy, as this results from seismic and radar interferometry data, it has been proved that prior to the Athens EQ two faults were activated. The objective of this paper is to support the hypothesis that the analyzed two kHz EM anomalies were sourced from the two faults. We show that the information and entropy content included in these two anomalies are quantitatively consistent with the seismic energy released, as it results from radar interferometry (Kontoes et al., 2000), as well as from seismic data analysis (Eftaxias et al., 2001), during the subsequent activation of two faults. The analysis is performed in terms of Fisher information and approximate entropy. We also refer to the already presented (Eftaxias et al., 2001) corresponding analysis for the EM energy for completeness. Finally, we discuss a series of arguments suggesting that the detection of the analyzed kHz EM activity was part of a series of findings from different disciplines indicating different mechanisms and different phases of the EQ preparation process, further supporting its precursor nature.

## 2　　Introduction to Fisher information and approximate entropy

In this section we briefly provide the basic background knowledge and some useful formulae concerning the Fisher information, and approximate entropy.

### 2.1　　Fisher information

Fisher information provides a powerful tool for the investigation of complex and non-stationary signals (Martin et al., 1999; Potirakis et al., 2012, and references therein). It has been used as a metric of the level of disorder of a system or phenomenon. It behaves inversely to entropy, i.e., increased order is characterized by decreased entropy, while increased Fisher information. It has been employed to study several geophysical and environmental phenomena, divulging its ability to describe the complexity of a system, e.g., (Balasco et al., 2008; Telesca et al., 2010), and suggesting its use to identify reliable precursors of critical events (Potirakis et al., 2012, and references therein).

In the case of a discrete measured variable $s_k = s(t_k)$, with $t_k = kT$, $k = 1,2,...,K$, and $T$ being the sampling period, one can define a set of $N$ disjoint but adjacent intervals (bins) covering the whole range of values between the minimum and maximum observed values of the time-series $\{s_k\}$, denoted as $\{x_n\}$, $n = 1,2,...,N$. The corresponding probability for a sample of the time-series to belong to the $n-$th interval can be denoted as $p(x_n)$. Then, Fisher information in its discrete form can be expressed (Martin et al., 1999) as:

$$I_x = \sum_{n=1}^{N-1} \frac{\left[p(x_{n+1}) - p(x_n)\right]^2}{p(x_n)}. \tag{1}$$

The discrete probability distribution $p(x_n)$ corresponds to the specific values of the unknown underlying probability density function at the centre values of the intervals $\{x_n\}$, which are not necessarily of equal length. The probability density function is usually approximated by the histogram, or by the kernel density estimator technique, employing different kernel functions like Gaussian kernel, or Epanechnikov kernel (Potirakis et al., 2012, and references therein).

## 2.2    Approximate entropy

Approximate entropy (ApEn) was introduced by Pincus (Pincus, 1991) as a measure of complexity or regularity that is applicable to noisy, medium-sized datasets. Since its introduction, ApEn has been widely applied to a variety of time-series of physiological and physical systems and has shown its superiority to most complexity measures such as fractal dimension, Kolmogorov–Sinai entropy, and spectral entropy (Karamanos et al., 2006).

For a time-series of a discrete measured variable $s_k = s(t_k)$, with $t_k = kT$, $k = 1, 2, ..., N$, and $T$ being the sampling period, we can define $N - m + 1$ vectors each one consisting of $m$ consecutive samples of this time series as

$$X_i^m = \{s_i, s_{i+1}, ..., s_{i+m-1}\}, \; i = 1, 2, ..., N - m + 1. \tag{2}$$

The main idea is to consider a window of length $m$ running through the time-series and forming the corresponding vectors $X_i^m$. The similarity between the formed vectors is used as a measure of the organization degree of the time-series. A quantitative measure of this similarity, $C_i^m(r)$, is $(N - m + 1)^{-1}$ times the number of vectors $X_j^m$ within a distance $r$ from $X_i^m$. By calculating it for each $i \leq N - m + 1$ and then taking the mean value of the corresponding natural logarithms, $\phi^m(r)$,

$$\phi^m(r) = (N - m + 1)^{-1} \sum_{i=1}^{N-m+1} \ln C_i^m(r) \tag{3}$$

ApEn is defined as

$$\text{ApEn}(m, r) = \lim_{N \to \infty} [\phi^m(r) - \phi^{m+1}(r)]. \tag{4}$$

ApEn is a "regularity statistics" that quantifies the unpredictability of fluctuations in a time series. The presence of repetitive patterns of fluctuation in a time series renders it more predictable than a time series in which such patterns are absent. A time series containing many repetitive patterns has a relatively small ApEn; a less predictable (i.e., more complex) process has a higher ApEn.

## 3    Electromagnetic data analysis

A way to examine transient phenomena is to analyze the pre-seismic EM time series into a sequence of distinct time windows. The aim is to discover a clear difference of dynamical characteristic as the catastrophic event is approaching. All employed metrics (Fisher information, approximate entropy and EM energy) were computed here versus time, by dividing the acquired time-series in consecutive non-overlapping time-windows of 512 samples length and computing all metrics for each one of them. In the case of Fisher information and approximate entropy the first difference of the raw data time series was analyzed, in order to remove the non-stationarities of the first order (Telesca et al., 2011).

## 3.1    The Fisher information approximate Entropy and EM energy vs time

The Fisher information (Fig. 1c) and approximate entropy (Fig. 1d), associated with successive time-windows were calculated and their evolution with time was studied. As it can be observed by the outliers of Fig. 1c, the unsmoothed Fisher information time pattern presents quasi-spike-like behavior, due to the signal's abrupt transitions (Telesca et al., 2010). The ApEn was calculated for $m=2$ and $r=0.65 \cdot STD$, where $STD$ is the standard deviation of the analyzed time-series fragment, allowing fragments with different amplitudes to be compared.

In Fig 1d, the $1-\mathrm{ApEn}$ is depicted for convenience, considering that the background noise entropy values correspond to entropy equal to 1. In both Figs 1c and 1d, only the information and entropy values which exceed the level of information and entropy of the background noise are depicted, respectively. This was accomplished by calculating the background noise information and entropy level during a quiet period of the signal, and then "subtracting" them from the calculated Fisher information and ApEn values, respectively.

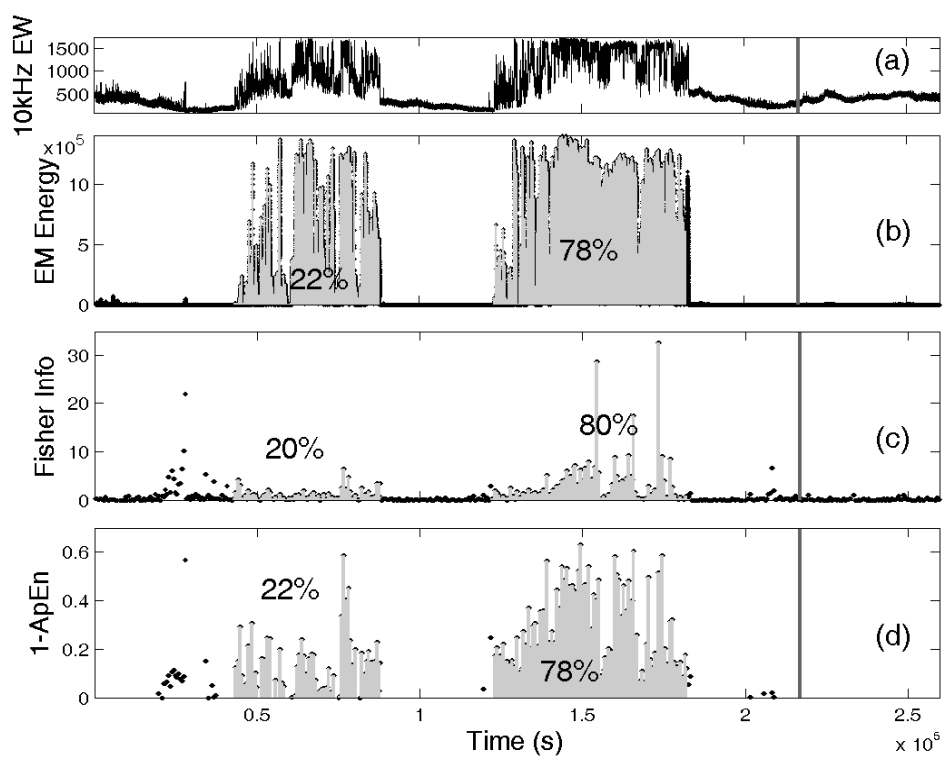

**Fig. 1.** (a) The two strong impulsive bursts in the tail of the recorded pre-seismic kHz EM emission (10kHz, East-West, magnetic field strength in arbitrary units) prior to Athens EQ (please refer to Fig. 1 in (Papadimitriou et al., 2008)). For the specific signal excerpt, the EM Energy (in arbitrary units) (b), the Fisher information (c) and approximate entropy (d) evolution with time are presented. The grey areas indicate the energy, information and 1-ApEn corresponding to the two bursts. The first (left) burst is responsible for the 22% of the EM energy, the 20% of the Fisher information, and the 22% of the ApEn, while the second (right) for the 78%, 80%, and 78% respectively. All graphs are time aligned for direct reference. The time of the EQ occurrence is indicated by the thick vertical grey line.

We observe that there is a significant increase of the Fisher information and a corresponding significant decrease of ApEn values during the two strong EM burst (the grey areas in Fig. 1c and Fig. 1d, respectively) and specifically during the second one. Fisher information and ApEn highlight them efficiently from the normal (background) behavior indicating their more ordered state (lower complexity).

If the Fisher information of the $j-\text{th}$ time window is denoted by $I_j$, then the information content of $w$ successive time-windows is considered here to be represented by the information sum $I_s = \sum_{j=n}^{n+w} I_j$, $j = n,\ldots, n+w$. In this context, we estimate that Fisher information content of the strong impulsive bursts in the tail of the pre-seismic EM emission of Fig. 1a, is unevenly distributed between the two bursts. Specifically, the first burst is responsible for the 20% of their total information content while the second for the remaining 80% (Fig. 1c). Similarly, the ApEn distribution between the two bursts was calculated to be 22% vs. 78% (Fig. 1d).

If the recorded precursory EM time-series amplitude is denoted by $A(t_i)$, the amplitude, $A_{fem}$, of a candidate "fracto-electromagnetic emission" is considered to be the difference $A_{fem}(t_i) = A(t_i) - A_{noise}$, where $A_{noise}$ is the maximum value of the EM recording during a quiet period, namely at a considerable time distance from the time of the EQ occurrence. The corresponding EM energy is estimated by the signal energy, $\varepsilon_i = \left(A_{fem}(t_i)\right)^2 \cdot \Delta t$, where $\Delta t$ is the sampling period (here 1 sec). We consider that a sequence of $k$ successively emerged "fracto-electromagnetic emissions" $A_{fem}(t_i)$, $i = m,\ldots, m+k$ represents the EM energy released, $E = \sum_{i=m}^{m+k} \varepsilon_i$, during the damage of a structure.

The temporal evolution of the EM energy of the analyzed time-series is shown in Fig. 1b. The energy release is also distributed so that the second burst contains ~80% of the total energy, as already indicated in (Eftaxias et al., 2001). To be more precise, the first burst is responsible for the 22% of their total energy release while the second for the remaining 78% (Fig. 1b).

## 4     Conclusions and discussion

We have shown that the ratios of the Fisher information, approximate entropy and EM energy, of the two kHz EM anomalies are consistent with the corresponding seismic energy ratio, as it results from radar interferometry (Kontoes et al., 2000), for the two faults activated during Athens EQ. For the first time a so strong relation between a pre-seismic signal and the ensuing fault activation appears in the literature simultaneously for energy, information and entropy content.

The application of multidisciplinary statistical analysis methods in terms of information and entropy (Papadimitriou et al., 2008; Potirakis et al., 2012) have been sensitively recognized and discriminated the candidate EM precursor under study. However, a stronger indication of the seismogenic origin of the signal would be (a) the presence of several universally holding scaling relations found in fracture and faulting processes, (b) the existence of "acceptable physical models to explain the existence of the precursor" (Cicerone et al., 2009), and (c) the compatibility to precursors from other disciplines. In our case all these three conditions are fulfilled. Indeed:

(a) The kHz EM emissions under study constitute a temporal fractal following the fBm model with roughness which is in harmony with the universal spatial roughness of fracture surfaces (Eftaxias et al., 2008). They behave as a "reduced image" of the regional seismicity, and a "magnified image" of laboratory seismicity (Papadimitriou et al., 2008).

(b) Both MHz and kHz signals are detected prior to large EQs. Moreover, the MHz radiation precedes the kHz one, as also observed in the case of the Athens EQ (Kapiris et al., 2004; Contoyiannis et al., 2005). Their generation has been supported by the following "two stage model of EQ generation": The MHz EM emission is thought to be due to the fracture of the highly heterogeneous system that surrounds the family of large high-strength entities distributed along the fault sustaining the system, while the kHz EM radiation is due to the fracture of the aforementioned large high-strength entities themselves (e.g., Kapiris et al., 2004; Contoyiannis et al., 2005). It has been proposed that the MHz signal results from a second order phase transition phenomenon, while the kHz signal results from a non-equilibrium instability phenomenon (Contoyiannis et al., 2005). Moreover, a new perspective to the kHz generation mechanism has been recently proposed (Minadakis, et al., 2012). In addition, Varotsos et al. (Varotsos et al., 2011) have reported that the occurrence time of a main shock is specified in advance by analyzing in "natural time" the seismicity subsequent to the initiation of the Seismic Electric Signals (SES) which are transient low frequency (≤1 Hz) electric signals that have been repeatedly recorded before earthquakes. This analysis identifies the time when the seismicity approaches the critical state: "the main shock was found empirically to follow usually within a few days up to one week". It is important to note that the MHz / kHz EM precursors are emerged from approximately a week up to a few hours before the EQ occurrence. We emphasize that, the MHz EM precursors can also be attributed to a phase transition of second order, as it happens for the seismicity preceding main shocks (Contoyiannis et al., 2005). In the frame of the proposed two stage model, the finally emerged kHz EM precursors indicate that the occurrence of the prepared EQ is unavoidable. This scheme, namely, the appearance of SES following by MHz - kHz EM precursory radiations, has been reported before the Athens EQ (Eftaxias et al., 2001).

(c) The appearance of signals from other disciplines indicated that an EQ preparation was in process around Athens as well. There was an acceleration of seismicity before the event, closely followed by the corresponding kHz EM emissions (Papadimitriou et al., 2008). Clear TIR (Thermal InfraRed) signals over the area around the Athens EQ epicenter were detected from satellites during the last days prior to the Athens EQ (Papadimitriou et al., 2008): after the 28-Aug-1999, a progressive increase (in extension and intensity) of the area affected started, reaching its maximum on 05-Sep-1999 (i.e. two days before the earthquake) and progressively dissipated after the event. The high-resolution topographic maps for measuring crustal strain accumulated over longer periods of time, obtained by radar interferometry (ERS-2 satellite), predicted the activation of two faults during Athens EQ (Kontoes et al., 2000).

Therefore, the probability for the analyzed signal to be a preseismic one seems to be increased. Not only because of the quantitative result that its energy, information and entropy contents are consistent with the seismic energy release during the subsequent activation of two faults, but also because of the presence of preseismic activity indications from other disciplines and the specific sequence of the emerged signals, corresponding to different generation mechanisms and different stages of EQ preparation process (SES, MHz, kHz).